\documentclass[twocolumn,11pt]{revtex4}
\usepackage{graphicx,epsf}
\usepackage{color}
\usepackage{dcolumn}
\usepackage{mathrsfs}

\usepackage{subfig}
			
\begin{document}

\title{Fast excitation of geodesic acoustic mode by   energetic particle beams}

\author{Jintao Cao$^1$, Zhiyong Qiu$^2$\footnote{E-mail: zqiu@zju.edu.cn }  and Fulvio Zonca$^{3,2}$ }

\affiliation{
$^1$Institute of Physics, Chinese Academy of Science, Beijing, P.R.C.\\
$^2$Inst.   Fusion Theory and Simulation, Zhejiang Univ., Hangzhou, P.R.C.\\
$^3$ ENEA C. R. Frascati, C. P. 65-00044 Frascati, Italy
}

\begin{abstract}
A new mechanism for Geodesic acoustic mode (GAM) excitation by a not fully slowed down energetic particle (EP) beam is analysed to explain  experimental observations in Large Helical Device. It is shown that  the positive velocity space  gradient near the lower-energy end of the EP distribution function can strongly drive GAM unstable. The new features of this EP-induced GAM  (EGAM) are: 1.  no instability  threshold in the pitch angle; 2. the EGAM frequency can be higher than the local GAM frequency; and  3. the instability  growth rate is much larger than that driven by a fully slowed down EP beam.
\end{abstract}

\maketitle

Geodesic acoustic modes (GAM) \cite{NWinsorPoF1968,FZoncaEPL2008} are finite frequency components of zonal structures (ZS), with a predominantly $n=0/m=0$ scalar potential and   $n=0/m=\pm1$ up-down asymmetric density perturbation. GAMs have been intensively studied in the past two decades due to their potential roles in regulating symmetry-breaking microscopic turbulence \cite{TSHahmPoP1999, ZLinScience1998,ZQiuPoP2014} and the associated wave-induced transport. Further to this wave-wave couplings, resonant wave-particle interactions with energetic particles (EP), i.e., the so called energetic-particle-induced GAM (EGAM),  also attracted significant attention \cite{RNazikianPRL2008,GFuPRL2008}. Due to their ``zonal" mode structures, GAM cannot be excited by expansion free energy associated with pressure gradients, and velocity space anisotropy is needed for EGAM instability. In previous works, a slowing-down EP distribution  with a localized pitch angle \cite{GFuPRL2008,ZQiuPPCF2010,HWangPRL2013} was adopted to study the EGAM excitation by neutral beam injection (NBI), considering that EP slowing down due to collisions with electrons is much faster than pitch angle scattering due to ion collisions. It was found that, for the fully slowed down EPs, only EPs with pitch angle $\Lambda B_0>2/5$ are destabilizing \cite{ZQiuPPCF2010},  as later confirmed by numerical simulation \cite{HWangPoP2013}. Here, $\Lambda=\mu/E$ is the pitch angle, $\mu=v^2_{\perp}/(2B)$ is the magentic moment, $B_0$ is the magnetic field at magnetic axis and $E=v^2/2$.
Berk and Zhou \cite{HBerkNF2010}, meanwhile, proposed that the sharp gradient in pitch angle induced by NBI prompt loss  could lead to   fast excitation of EGAM. GAM excitation by a bump-on-tail ion distribution function is addressed in Refs. \citenum{DZarzosoPRL2013,JBaptistePoP2014}, for the case where EP source is the high-energy tail in ion distribution function induced by radial frequency (RF) heating; and possible interaction between EGAM and drift wave turbulence is discussed \cite{DZarzosoPRL2013,ZQiuEPS2014,RDumontPPCF2013}.

A recent paper by Ido et al \cite{TIdoNF2015} presented experimental  evidences  in the Large Helical Device (LHD), where EGAM is observed during tangential neutral beam injection, and  some peculiarities appears in the comparison with  theoretical predictions. The EGAM is excited before the injected NBI is fully slowed down, and the EP birth energy ($\sim170KeV$) is much larger than that predicted for wave-particle resonance \cite{ZQiuPPCF2010}; thus, earlier theories based on a fully slowed down NBI \cite{GFuPRL2008,ZQiuPPCF2010} cannot be directly applied here.  In this work, we explain why EGAM is excited by a not fully slowed down EP beam, with the EP distribution function being a function of time. As we will show later, the instability drive comes from the positive velocity space gradient in the low-energy end of the EP distribution function,    similar to that of a bump-on-tail distribution. The EGAM excited by such an EP beam can be applied to explain the experimental observations in LHD \cite{TIdoNF2015}.
For the sake of clarity , we focus only on the local dispersion relation of EGAM, while neglecting system nonuniformity and  higher order effects, such as finite Larmor radius. Modulation of EP distribution function due to excitation of EGAM is not taken into account either.

The dispersion relation of the $n=0$ electrostatic  EGAM can be derived from quasi-neutrality condition,
\begin{eqnarray}
\sum_s\left\langle\frac{Q}{m}\frac{\partial F_0}{\partial E}\delta \phi+J_0(k_r\rho_L)\delta H_g\right\rangle_s=0,\label{QNcondition}
\end{eqnarray}
with  $\delta H_g$ being the nonadiabatic component of the perturbed particle distribution function
\begin{eqnarray}
\delta f_s=\frac{Q_s}{m_s}\frac{\partial F_{0,s}}{\partial E}\delta \phi+\exp\left\{i\frac{mc}{QB^2}\mathbf{k}\times\mathbf{B}\cdot\mathbf{v}\right\}\delta H_g,\nonumber
\end{eqnarray}
and $\delta H_g$ is derived from  linear gyrokinetic equation\cite{PRutherfordPoF1968,JTaylorPP1968,EFriemanPoF1982}
\begin{eqnarray}
&&\left(\omega-\omega_d+i\omega_{tr}\frac{d}{d\theta}\right)\delta H_g\nonumber\\
&=&-\frac{Q_s}{m_s}\frac{\partial F_{0,s}}{\partial E}J_0(k_r\rho_L)\delta\phi.
\end{eqnarray}
Here, the subscripts $s=e,i,h$ representing electrons, ions and EPs, respectively; and $Q$ is the electric charge. Furthermore,  $J_0(k_r\rho_L)$ is Bessel function of the zero-index, with $\rho_L=mcv_{\perp}/(BQ)$ being the Larmor radius, $\omega_{tr}=v_{\parallel}/(qR_0)$ is the transit frequency,  $\omega_d=-k_r(v^2_{\perp}+2
v^2_{\parallel})/(2 \Omega R_0)\sin\theta$ is the magnetic drift frequency; and other notations are standard.

The thermal plasma response to GAM  has already been derived in several earlier works, and we shall not provide the details here. EP response can be derived by transforming into drift orbit center coordinates.  Since, in the scenario considered here, EGAM is excited before the EPs are fully slowed down, we assume the following distribution for EPs:
\begin{eqnarray}
F_{0h}=\frac{c_0H(E_b-E)H(E-E_L)}{E^{3/2}+E^{3/2}_{crit}}\delta(\Lambda-\Lambda_0).
\end{eqnarray}
Here, $c_0$ is related with NBI flux and slowing-down time,  $H$ is the Heaviside step function,  $E_b$ is the EP birth energy, $E_L\simeq E_b\exp(-2\gamma_c t)$ is the lower end of the distribution function, $\gamma_c$ is the slowing-down rate,  and we typically have $E_b\gg E_L\gg E_{crit}$.
Assuming small EP drift orbit  normalized to EGAM wave length and keeping only $l=\pm1$ transit harmonics, one then obtains the following dispersion relation from  the surface averaged quasi-neutrality condition
\begin{equation}
-\frac{e}{m}n_{c}k^2_r\frac{1}{\Omega^2_i}\left(1-\frac{\omega^2_G}{\omega^2}\right)\overline{\delta\phi}+\overline{\delta
n_h}  = 0 \label{qn},
\end{equation}
where $\omega_G$ is the GAM frequency, $A=ecQ^2k^2_r\overline{\delta\phi_G}/(\sqrt{2}B_0\Omega_i)$ and the perturbed EP density
\begin{eqnarray}
\overline{\delta n_{h}}&\simeq&2\pi
B\sum_{\sigma=\pm1}\int\frac{Ed\Lambda
dE}{|v_{\parallel}|}\overline{\delta H_h}\nonumber\\
&=&2A\int\frac{(2-\Lambda B)^2}{(1-\Lambda
B)^{1/2}}\frac{BdEd\Lambda E^{5/2}\partial_EF_{0h}}{2E(1-\Lambda
B)-\omega^2q^2R^2_0}.\nonumber\\
\end{eqnarray}

The dispersion relation can then be derived as
\begin{eqnarray}
&&-1+\frac{\omega^2_G}{\omega^2}+\frac{\pi}{\sqrt{2}}B_0q^2\frac{c_0}{n_0}\times\nonumber\\
&&\left[C\left(\ln(1-\omega^2/\omega^2_b)-\ln(1-\omega^2/\omega^2_L)\right)\right.\nonumber\\
&&\left.+D\left(\frac{1}{1-\omega^2_b/\omega^2}-\frac{1}{1-\omega^2_L/\omega^2}\right)\right]=0.
\end{eqnarray}
Here, $C=(2-\Lambda_0B_0)(5\Lambda_0B_0-2)/(2(1-\Lambda_0B_0)^{5/2})$, $D=\Lambda_0B_0(2-\Lambda_0B_0)^2/(1-\Lambda_0B_0)^{5/2}$, and $\omega_b$ and $\omega_L$ are the transit frequency defined at $E_b$ and $E_L$, respectively. Note that $\omega_L$ is a slowly varying function of $t$.

Note that there are two different singularities in the dispersion relation; i.e., the logarithmic terms (the first ones in the square bracket) typical of a slowing down distribution, and the simple poles (the second ones in the square bracket) from the integration limits. As noted in Ref. \citenum{ZQiuPPCF2010} , the simple pole at $1-\omega_b^2/\omega^2=0$  only modulate theEGAM  real frequency, but it does not contribute to mode excitation. The  logarithmic terms, on the other hand, are destabilizing for $\Lambda_0B_0>2/5$; thus, there is a threshold,  $\Lambda_0B_0>2/5$,   for EGAM excitation by the fully slowed down EP beam case \cite{ZQiuPPCF2010}. Meanwhile, for the not fully slowed down distribution function considered here, the simple pole at $1-\omega^2_L/\omega^2=0$ is also destabilizing;  thus, there is no threshold in the pitch angle $\Lambda_0B_0$ for EGAM excitation by the EPs.

\begin{figure}
\includegraphics[width=9cm]{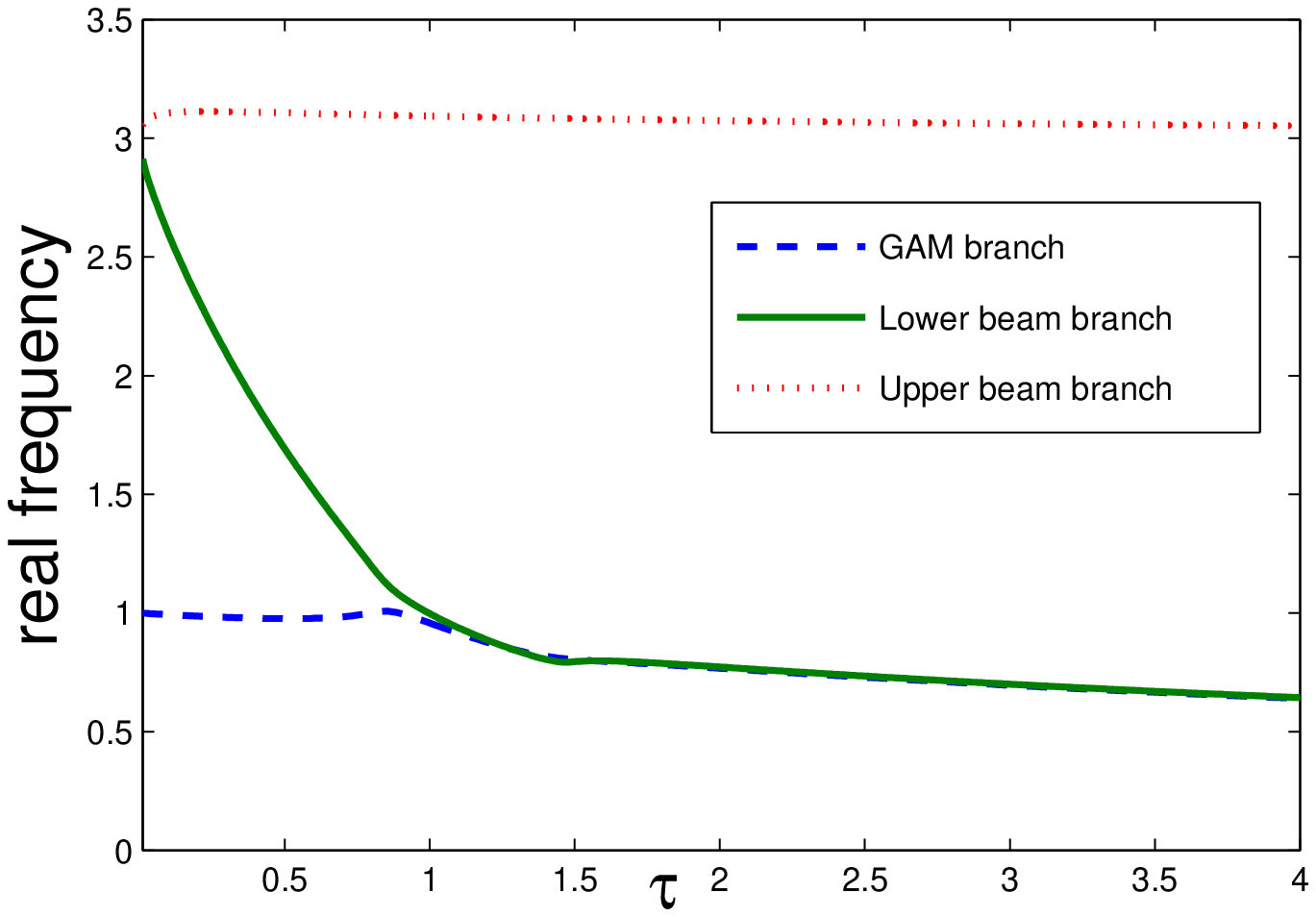}
\caption{Real frequency for $\Lambda_0B_0>2/5$}\label{fig:RF_LPA}
\end{figure}

\begin{figure}
\includegraphics[width=9cm]{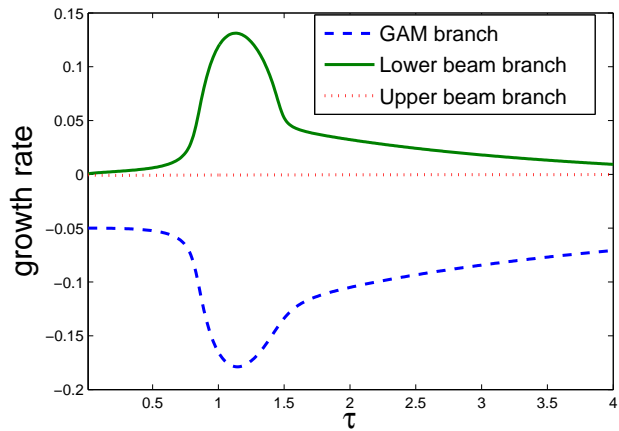}
\caption{Growth rate for $\Lambda_0B_0>2/5$}\label{fig:GR_LPA}
\end{figure}

The dispersion relation can be solved numerically as a function of $\tau=\gamma_c t$, and yield the slow temporal evolution of the excited EGAM due to the slowing down of the EP beam. There are three branches;  a GAM branch with $\omega_r\simeq \omega_G$, a lower beam branch (LBB) with $\omega_r\simeq \omega_L(t)$; and an upper beam branch(UBB), with $\omega\simeq \omega_b$. Here, we take $\omega_G=1-0.05i$ with the imaginary part being the Landau damping rate of GAM, $\omega_b=3\gg |\omega_G|$, and $\Lambda_0B_0>2/5$ such that the coefficient of the logarithmic term (i.e., $C$)  is positive. The real frequency and growth rate are shown, respectively, in Figs. \ref{fig:RF_LPA} and \ref{fig:GR_LPA}. The horizontal axis is time in unites of $\gamma^{-1}_c$. We may see that, only LBB with the frequency $\omega\simeq\omega_L$ can be unstable, consistent with the fact that, in the present case of experimental interest for LHD experimental observations \cite{TIdoNF2015}, the UBB frequency is larger than the GAM frequency \cite{ZQiuPPCF2010}. The LBB is stable when $\omega_L\gg\omega_G$; as $\omega_L$ approaches $\omega_G$ with $t$, the growth rate of LBB will increase significantly ($t=0.8\sim1.2$), and then decays quickly as $E_L$ further decreases ($t=1.2\sim1.5$). Later on ($t>1.5$), the growth rate decays very slowly, due to the contribution of the destabilizing logarithmic term. So for $\Lambda_0B_0>2/5$, the mode can be interpreted as a double-beam plasma instability with the two singularities dominating at different times. As $\omega_L$ approaches $\omega_G$, the simple pole  dominates; however, as $\omega_L$ further decreases and becomes smaller than $\omega_G$ by a finite amount, the contribution from the simple pole is negligible, while the logarithmic term still contributes to destabilizing the EGAM \cite{ZQiuPPCF2010}.  The strong instability at $\omega_L(t)\simeq\omega_G$ may also provide an explanation for the fast growth of  EGAM observed experimentally.
We note also that, the frequency of the unstable LBB, can be significantly larger than $\omega_G$, as is shown in Fig. \ref{fig:RF_LPA}.  This may explain the higher-frequency branch of EGAM observed in LHD \cite{TIdoNF2015}. Another evidence of that  the present theory can be applied to explain the LHD experiment is that, in the discharges with electron cyclotron resonance heating (ECH), the dependence of the observed EGAM frequency on plasma temperature is weaker. This is because that the slowing down rate is lower in the ECH discharges, where collisional rate is lower due to higher electron temperature, such that the unstable mode is more a ``beam-branch"  with the frequency determined by the transit frequency of EPs \cite{ZQiuPPCF2010}.

\begin{figure}
\includegraphics[width=9cm]{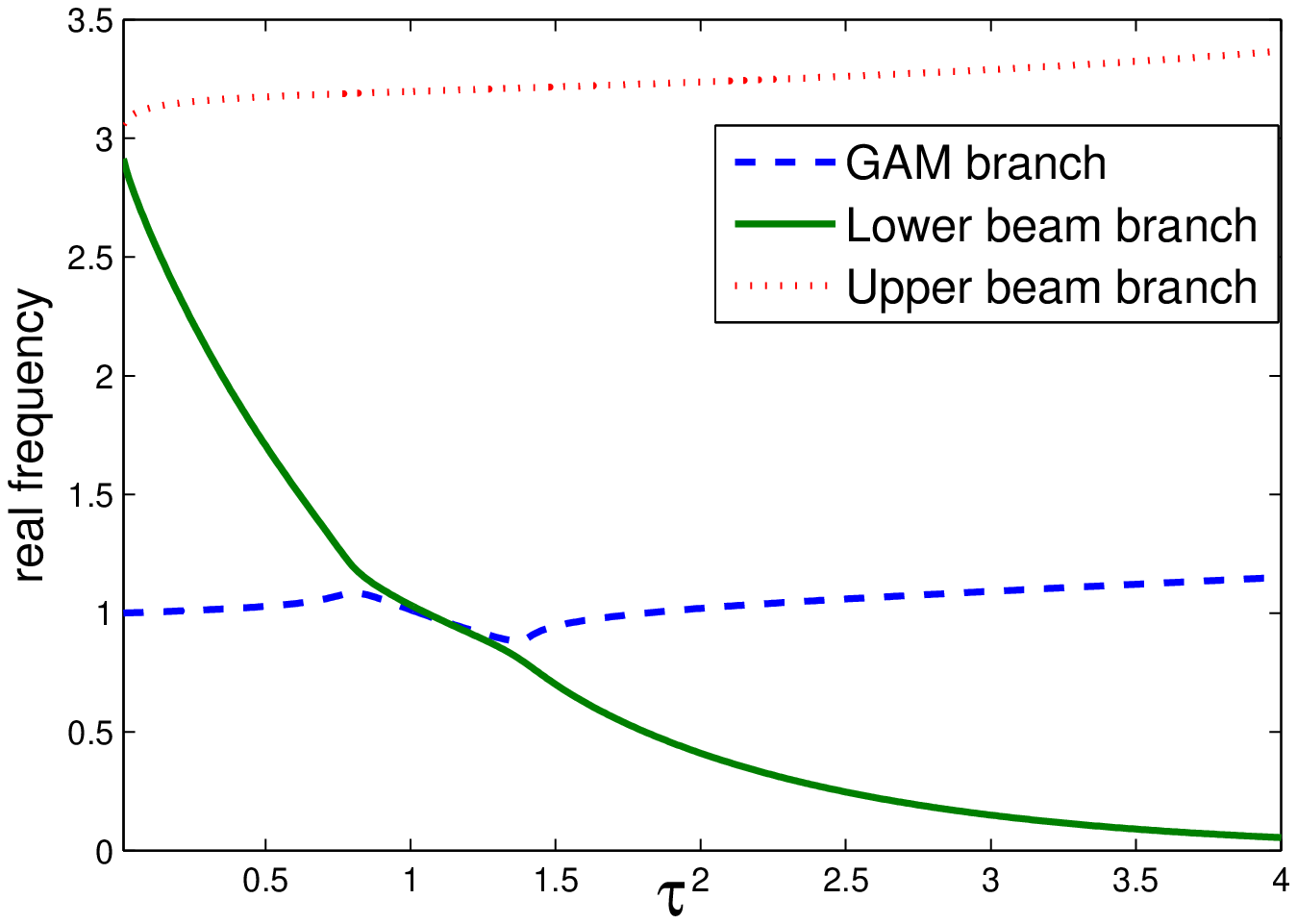}
\caption{Real frequency for $\Lambda_0B_0<2/5$}\label{fig:RF_SPA}
\end{figure}
\begin{figure}
\includegraphics[width=9cm]{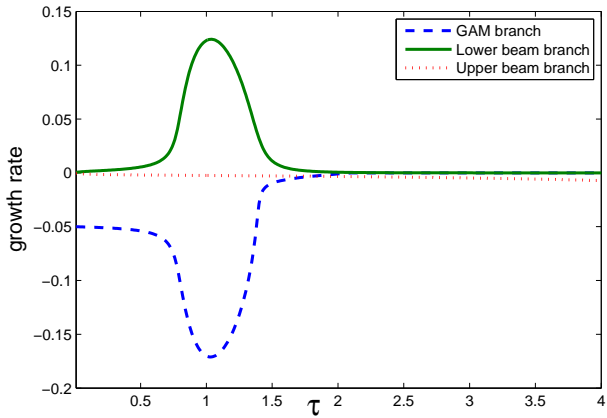}
\caption{Growth rate for $\Lambda_0B_0<2/5$}\label{fig:GR_SPA}
\end{figure}

The real frequency and growth rate for $\Lambda_0B_0<2/5$  are shown in Figs. \ref{fig:RF_SPA} and \ref{fig:GR_SPA}, respectively. In this case, the logarithmic term is stabilizing \cite{ZQiuPPCF2010}; thus,  the EGAM discussed here is similar to beam-plasma instability, which, however, has a double pole instead of the simple pole of the present case.  We can see that, the behaviors of the three branches are similar to the case with $\Lambda_0B_0>2/5$. However, when $\omega_L$ becomes smaller than $\omega_G$ by a finite amount, the growth rate of LBB decreases to zero as the contribution of the simple pole becomes vanishingly small, similar to that of a beam-plasma instability. But it is clearly shown that, due to the contribution from the positive gradient in the lower-energy end of the distribution function, EGAM can still be driven unstable here, unlike the case of the fully slowed-down EP beam.

In conclusion, we elucidate the mechanism for GAM excitation by a not fully slowed down EP beam, which can be applied to explain the experimental observations in Large Helical Device. There are two kinds of  singularities in the dispersion relation;   one is the logarithmic singularity typical of a slowing-down distribution function; and the other one is simple pole  from the low-  and high-energy  end of the distribution function. It is found that only the lower beam branch with the frequency determined by the low-energy limit of the distribution function is unstable, consistent with the fact that the upper beam branch frequency is larger than the GAM frequency as in the case of experimental interest. For $\Lambda_0B_0>2/5$, both singularities are destabilizing, with the simple pole dominating as $\omega_L\simeq\omega_G$;  and the contribution of the logarithmic singularity  taking over as $\omega_L$ becomes much smaller than $\omega_G$. On the other hand, for $\Lambda_0B_0<2/5$, only the simple pole is destabilizing,  and the behavior of EGAM is very similar to a beam-plasma instability.  In both cases, the real frequency of the unstable lower beam branch can be significantly higher than GAM frequency, providing an explanation for   experimental observations. Note that the real frequency of the unstable lower beam branch can be higher than local GAM frequency doesn't conflict with the argument we used to explain why the upper beam branch is always stable, since the simple pole at $\omega\simeq\omega_{b}$ is always stabilizing, so one needs  the unstable branch to have a frequency lower than local GAM frequency for the instability of the logarithmic term; while the lower beam branch is unstable at $\omega_{L}>\omega_G$ due to the contribution of the simple pole.
Another novel result of this work is that,  the EGAM drive is much stronger than that predicted in earlier theories \cite{GFuPRL2008,ZQiuPPCF2010} , since the destabilizing simple pole singularity obtained in the present case  is much stronger than the logarithmic singularity from a fully slowed down EP distribution function.

\section*{Acknowledgments}
This work is supported by   the ITER-CN under Grants Nos.
2013GB104004, 2013GB112011  and   2013GB111004,
the National Science Foundation of China under grant Nos. 11235009, 11205132 and 11575157,
Fundamental Research Fund for Chinese Central Universities and   EUROfusion Consortium
under grant agreement No. 633053. Discussion with Prof. Liu Chen (Zhejiang Unviersity and University of California, Irvine) is kindly acknowledged.

\end{document}